\documentclass[12pt]{revtex4}
\usepackage{amsmath, amsthm, amssymb}
\usepackage{graphics}
\usepackage{graphicx}
\usepackage{epstopdf}
\usepackage{subfigure}
\usepackage{longtable}
\usepackage{url}
\usepackage{bm}
\usepackage{float}

\begin{document}
\noindent The following article has been accepted by Applied Physics Letters.  After it is published, it will be found online at http://apl.aip.org

\title{Spontaneous emission and collection efficiency enhancement of single emitters in diamond via plasmonic cavities and gratings}

\author{Jennifer T. Choy$^1$, Irfan Bulu$^1$, Birgit J. M. Hausmann$^1$, Erika Janitz$^{1,2}$, I-Chun Huang$^1$, and Marko Lon\v{c}ar$^1$}

\affiliation{$^1$School of Engineering and Applied Sciences, Harvard University, Cambridge, MA, USA, 02138}

\affiliation{$^2$Institute for Quantum Computing, University of Waterloo, Waterloo, ON, Canada, N2L 3G1}

\maketitle
\section*{Abstract} We demonstrate an approach, based on plasmonic apertures and gratings, to enhance the radiative decay rate of single NV centers in diamond, while simultaneously improving their collection efficiency.  Our structures are based on metallic resonators formed by surrounding sub-wavelength diamond nanoposts with a silver film, which can enhance the spontaneous emission rate of an embedded NV center.  However, the collection efficiency of emitted photons remains low due to losses to surface plasmons and reflections at the diamond-air interface.  In this work, we mitigate photon losses into these channels by incorporating grating structures into the plasmonic cavity system.

\section*{Main Text}
Diamond is a promising material for integrated quantum photonics due to its excellent material properties and the optical response of its defect centers, which can include stable single photon emission at room temperature.  In particular, the nitrogen-vacancy (NV) center provides optical addressability and readout of electron and nuclear spin states that are associated with long coherence times~\cite{Jelezko_PSS}, which make it ideal for applications such as quantum information processing, quantum communication~\cite{Childress2013}, and magnetometry~\cite{Hong2013}.  For these applications, it is advantageous to embed color centers into diamond-based nanophotonic devices to control the efficiencies of single photon production and collection~\cite{Hausmann_PSS}.  

We have previously presented the ``diamond-silver aperture,'' consisting of a diamond nanopost (of radius $r_d\approx50$ nm and height $h_d\approx220$ nm) covered in silver (Ag)~\cite{Bulu2011, Choy2011}, which supports guided modes that become more tightly confined in the dielectric with decreasing $r_d$.  Reflections between the facets of the aperture form resonances that can be tailored by changing $r_d$ and $h_d$ and used to modify the spontaneous emission rate (SE) of an interacting NV center in the nanopost.  The hybrid structure can be implemented by a combination of ion implantation and top-down nanofabrication, and the coupling between the emitter and resonator has yielded a six-fold Purcell enhancement of the NV center SE compared to emission in the bulk~\cite{Choy2011}.  While the fluorescence intensities from NV centers in the diamond-Ag apertures have increased from those in the bulk due to faster radiative decay rates, only about 3\% of the emitted photons are collected, due to coupling of the emission to surface plasmons and total internal reflection (TIR) at the diamond-air interface.  Therefore, it is desirable to engineer the directionality of the emission via periodic corrugations in the metal film~\cite{Lezec2002, Carretero2011} which coherently scatter surface plasmons, thereby increasing the fraction of optical power going into the 25$^\circ$ critical cone set by TIR.  In this work, we experimentally demonstrate a platform to integrate the apertures with plasmonic gratings to improve the collection efficiency of the emitted single photons from cavity-enhanced NV centers.

\begin{figure}[H]
  \centering
\includegraphics[width=\columnwidth]{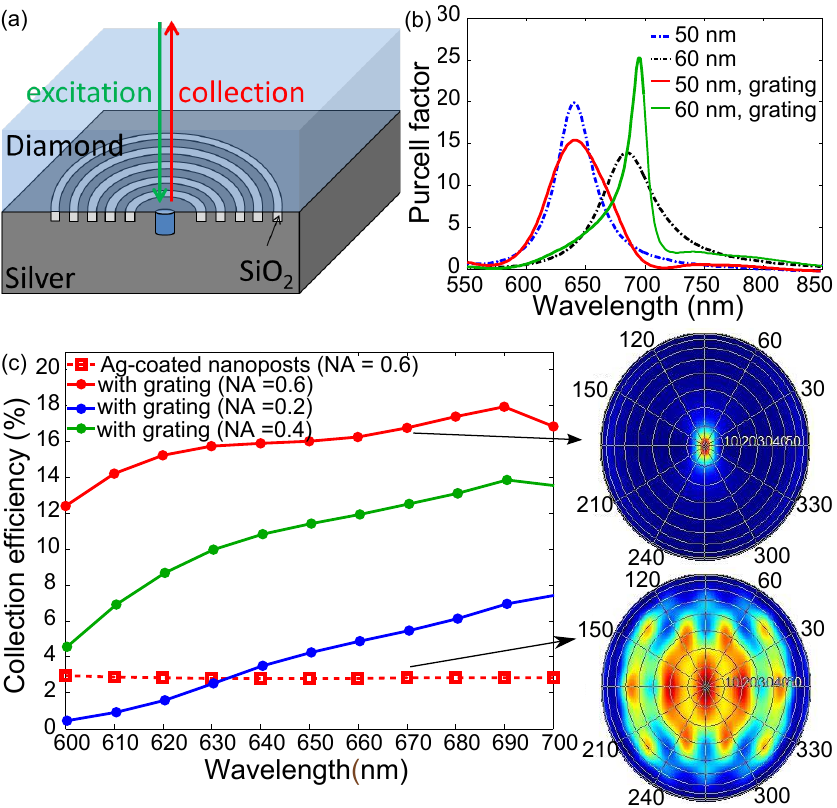}
\caption{(a) Schematic of the diamond-Ag grating structure.  (b) Calculated Purcell factor for Ag-coated nanoposts with and without grating.  (c) Plot of collection efficiency as a function of wavelength for the non-grating and grating cases under different collection optics.  Our microscope objective has a N.A. of 0.6.  Insets: simulated angular distributions in diamond for the power emitted from (top) a diamond-Ag aperture surrounded by grating and (bottom) the same-radius aperture without grating at a wavelength of 670 nm.}
\label{figure1}
\end{figure}

Similar approaches to provide more efficient extraction of emitted light and to manipulate the angular pattern of the radiation have been proposed~\cite{Maksymov2010} and demonstrated with quantum dots~\cite{Curto2011, Livneh2011, Jun2011}, fluorescent molecules~\cite{Taminiau2008, Aouani2011}, and NV centers in diamond nanocrystals~\cite{Wolters2012, Mollet2012}.  The geometry presented here is based on emitters in the bulk crystal, thus avoiding the placement techniques generally associated with nanocrystal platforms.  We add concentric metallic rings around the diamond-Ag nano-cavity, such that the final structure (Fig.~\ref{figure1}a) consists of periodic grooves in the Ag film filled with a dielectric material around a central aperture.  For the dielectric filler, we have selected silicon dioxide (SiO$_2$) rather than the more obvious choice of diamond so that the gratings can be controlled independently from the diamond nanoposts.  Additionally, since our sample fabrication involves using blanket ion implantation to generate a layer of NV centers, molding the grooves with SiO$_2$ can avoid contribution to the single photon emission by extraneous fluorescence from embedded emitters in the gratings.

We performed 3D finite-difference-time-domain (FDTD) simulations to model our structures.  The diamond aperture has $r_d$ between 50 nm and 60 nm and $h_d$ = 220 nm, with a radially polarized dipole optimally placed in the field maximum of the cavity.  The Purcell factor~\cite{Purcell}, which is the SE rate enhancement experienced by the dipole in cavity, is calculated~\cite{Bulu2011} and shown in Fig.~\ref{figure1}b.  The resonant wavelengths range from 640 nm to 680 nm, which overlap well with the zero phonon line (637 nm) and the peak emission (around 670 nm) of the NV center fluorescence, respectively.  

The general grating design is similar to that described in~\cite{Carretero2011}, and consists of five grooves, each with the same width $w$ = 80 nm and height $h_g$ = 100 nm.  The distance between the first groove and center of the structure is $a$ = 350 nm and the periodicity of the subsequent grooves is $b$ = 280 nm.  These parameters were selected to overlap the resonant response of the gratings with that of the center aperture, and it can be seen in Fig.~\ref{figure1}b that the addition of grating has led to an increase in the SE enhancement when the resonant responses are matched, while otherwise the grating induces only small shifts in the cavity resonance wavelength for the range of post radii we studied.  The interference between the cavity and grating resonances has also led to a Fano lineshape.  The addition of the plasmonic corrugations increases the overall collection efficiency up to 16\% over a broad range of wavelengths around the resonance, which represents a 5-fold improvement over the bare aperture case (around 3\%).  This is shown in Fig.~\ref{figure1}c, and was calculated by projecting the electric fields obtained above the device into the far-field, applying the Fresnel coefficients and Snell's law~\cite{Hecht} to take into account the diamond-air interface, and finally dividing the integrated power emitted into the 37$^\circ$ cone set by our collection objective (with numerical aperture N.A. of 0.6) by the total power emitted by the dipole.  We have also plotted the collection efficiency under collection with different N.A., and an efficiency of around 10\% is still possible with a N.A. of 0.2-0.4, suggesting that the structure might be compatible with direct collection by optical fibers.  The improvement in collection efficiency is accompanied by modification of the far field profile of the radiation, in which emission is strongly concentrated at small angles (Fig.~\ref{figure1}c).  

\begin{figure}[H]
  \centering
\includegraphics[width=\columnwidth]{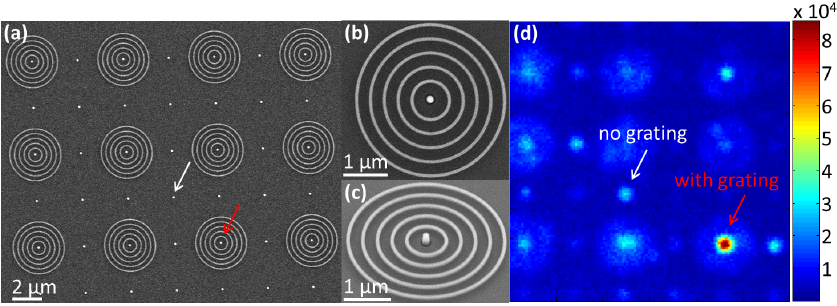}
\caption{(a)-(c) SEM (taken before Ag deposition) and (d) confocal scan images (taken after sample has been embedded in Ag) of arrays of diamond nanoposts (white arrow) in which alternating posts are surrounded by SiO$_2$ rings (red arrow).}
\label{figure2}
\end{figure}

We fabricated arrays of diamond nanoposts of radii from 50 nm to 70 nm and height around 240 nm in an electronic grade diamond that was implanted with a shallow layer of NV centers 90 nm below the surface, using electron beam lithography and reactive ion etching in an O$_2$ plasma~\cite{Hausmann_PSS}.  To realize the plasmonic corrugations, we defined rings (with grating periods $a$ from 350 nm to 360 nm and $b$ from 280 nm to 300 nm) of flowable oxide around the fabricated nanoposts in a second electron beam lithography step using a negative electron beam resist (hydrogen silsesquioxane).  Scanning electron microscope (SEM) images indicate that the gratings and pre-exsting structures are well aligned (Fig.~\ref{figure2}a-c).  The images shown here are taken of a different set of samples than the ones optically characterized; this is to avoid introducing contamination (such as carbon deposition) to the devices during the imaging process that can lead to unwanted fluorescence.  To further ensure the cleanliness of the surface, we cleaned the diamond sample in a Piranha solution (3:1 H$_2$SO$_4$:H$_2$O$_2$) and annealed the sample at 465$^\circ$C in an O$_2$ purged environment.  The final structure was realized by capping device with sputtered Ag.

\begin{figure}[H]
  \centering
\includegraphics[width=\columnwidth]{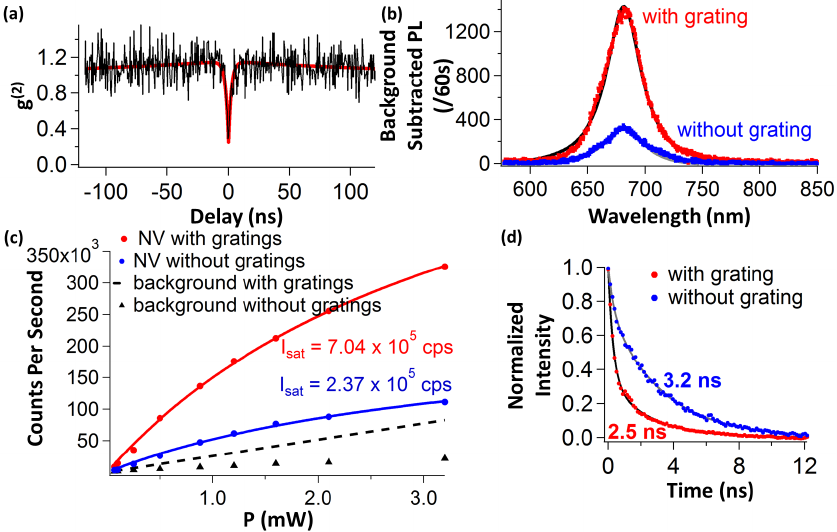}
\caption{(a) g$^{\mathrm{(2)}}$ plot of a diamond-Ag nanopost with grating.  The black line indicates raw data, while the red line is the fit.  Comparison of (b) photoluminescence (PL) spectra, (c) saturation intensities (the background-subtracted fluorescence, and the background count rates are shown), and (d) fluorescence time decay traces of the grating and non-grating cases.  The solid lines in (d) indicate fits to an exponential model.}
\label{figure3}
\end{figure}

As mentioned earlier, we characterized the diamond-Ag apertures by exciting the sample with a continuous wave pump laser at 532 nm and collecting the NV center emission (filtered by a bandpass filter from 650-800 nm) through the bulk diamond crystal in a home-built confocal microscope with a 0.6 N.A. microscope objective~\cite{Choy2011}.  The fluorescence was focused onto one end of a single-mode, 2$\times$2 fiber coupler, with each output connected to an avalanche photodiode (APD) for photodetection.  A confocal scan image of the sample (Fig.~\ref{figure2}d) from one of the APD outputs shows that emission from posts surrounded by gratings are brighter than those without gratings under the same pump power (around 1 mW).  To confirm single photon character from individual NV centers in the structures, the photon arrival times at the APDs were recorded and plotted in a histogram of coincidence counts as a function of time delay.  After normalization, we obtain the second order autocorrelation function, g$^{(2)}$($\tau$) (Fig.~\ref{figure3}a).  For the two devices indicated with arrows in Fig~\ref{figure2}d, g$^{(2)}(0)<0.5$, demonstrating that they each contain a single emitter.   These nanoposts have the same post radius of 55 nm, so that their resonator characteristics should be very similar (as can be seen by the background-subtracted photoluminescence spectra shown in Fig.~\ref{figure3}b), thus allowing us to compare the effect of the grating.

The fluorescence intensities, $I$, of the nano-posts were measured under different pump powers $P$ (Fig.~\ref{figure3}c).  The background contribution to the fluorescence was inferred directly from g$^{(2)}$ measured at various $P$~\cite{Beveratos2001, supplementary}.  The background subtracted count rate is the NV fluorescence intensity and is fitted to the saturation model, $I(P) = \frac{I_{sat}}{1+P_{sat}/P}$, where $I_{sat}$ and $P_{sat}$ are the saturation intensity and power, respectively.  $I_{sat}$ of the post with grating is 704$\pm$38 kcps while that for the bare cavity is 237$\pm$17 kcps.  $P_{sat}$ are comparable in the two cases.  

Contributions to increased $I_{sat}$ can be attributed to improvement in collection efficiency and/or Purcell enhancement.  The latter can be inferred from the emitter lifetime, which was measured using excitation by green pulses at a 76 MHz repetition rate (Fig.~\ref{figure3}d).  The fluorescence intensity time traces were fitted to an exponential decay model, and the resulting time constants are 3.2 ns and 2.5 ns for the non-grating and grating cases, respectively.  The corresponding Purcell factors, calculated by comparing the NV lifetime in cavity with that in the bulk (around 16 ns in 90-nm implanted NV centers), are roughly 6.  The discrepancies between the observed Purcell enhancements and the predicted values can be attributed to geometric effects (most notably the tapering in the sidewalls of the nanoposts from the reactive ion etching process), straggle in the implantation depth, emitter spectrum, and misalignment between the polarization angle of the dipole moment and the [100] diamond crystal plane.  In particular, we calculated that the peak emission enhancement is reduced by about 40\% from the straight sidewall case for a 85$^\circ$ sidewall angle~\cite{supplementary}, which matches well with experimentally-obtained values.  Taking Purcell enhancements into account, the improvement in collection efficiency provided by the grating is around 2.3.

\begin{figure}[H]
  \centering
\includegraphics[width=\columnwidth]{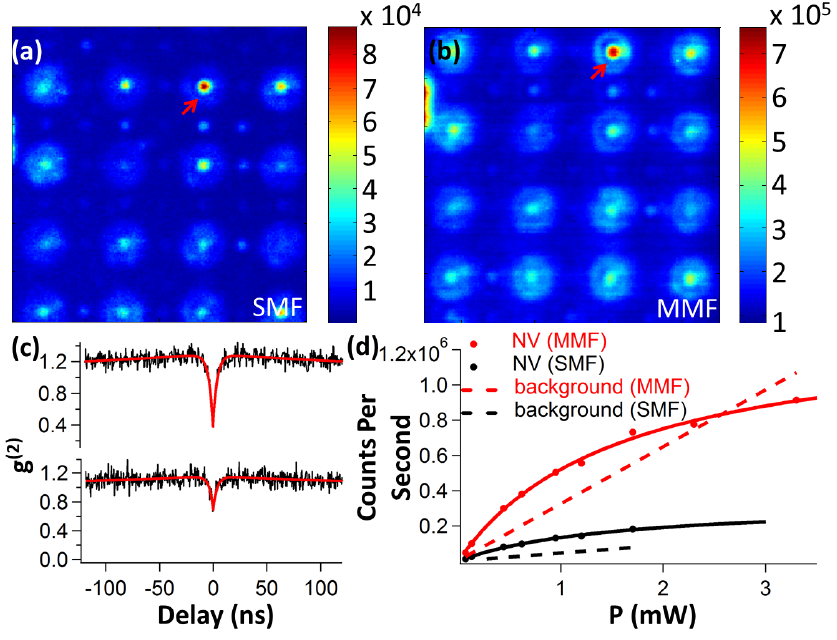}
\caption{Confocal scans of the same area of the sample as imaged using a (a) SMF and (b) MMF at the same power.  (c) g$^{\mathrm{(2)}}$ plot of the device (indicated by the red arrows) under collection by a SMF (top) and MMF (bottom), taken at comparable powers.  (d) Saturation curves under collection by a SMF and MMF.  $I_{sat}$ are $3.37\times10^5$ cps and $1.35\times10^6$ cps, respectively.  $P_{sat}$ are around 1.5 mW for both fits.}
\label{figure4}
\end{figure}

Finally, we consider the effect of pinhole filtering on the overall collection efficiency of the setup.  While the footprint of our grating devices spans over $5\ \mu m$, the measured spot size on our confocal microscope has a full-width half-maximum of 610 nm.  Therefore, some of the light scattered by the grating is not collected by our confocal system.   To increase the area of collection, we replaced the single-mode fiber (SMF), which has a core size of around 4.5 $\mu m$, with a multi-mode fiber (MMF) with a core size of 62.5 $\mu$m.  Since the role of the pinhole is to reduce the area and depth of collection, enlarging it would increase the area being probed in a way that is proportional to the increase in its size.  However, we also sacrificed the filtering of unfocused light which led to an overall increase in background coming from the larger volume of diamond being probed.  Nonetheless, as shown in Fig.~\ref{figure4}a-b, the scan images taken with SMF and MMF show similar features, with increased collected count rates by several fold in the latter.  Due to the higher background, g$^{(2)}$(0) is slightly above 0.5 in the MMF case (Fig.~\ref{figure4}c).  To quantify the intensity increase in a grating device with confirmed single photon character, we compared its saturation behaviors under the two collection channels (Fig.~\ref{figure4}d).  After background subtraction, $I_{sat}$ is almost 4 times higher than that collected by a SMF (from 337 kcps to 1.35 Mcps), while $P_{sat}$ is mostly unchanged between the two measurements.  Photon count rates of the structures under MMF collection are thus comparable to those obtained from nanowires ($\sim500$ kcps)~\cite{Babinec2010} and solid-immersion lenses ($\sim2$ Mcps)~\cite{Schroder2011}.  Ultimately, to optimize both the collection of all scattered signal from the grating and spatial filtering of the background, a variable pinhole may be used.

We have experimentally demonstrated the integration of plasmonic gratings with single NV centers in diamond-Ag apertures.  With some improvements mentioned below, our system is potentially promising for shaping and collimating single photon beams on chip.  The observed improvement in collection efficiency is modest (a factor of 2.3) and is about half of the predicted value.  This could be attributed to deviation of the dipole from the field maximum of the cavity, which affects the coupling of emission to surface plasmons.  Additionally, the slight taper in the sidewall degrades the confinement and further reduces the coupling to propagating surface plasmons. 

Overall, the device yield is poor, with only about 4 working grating devices (qualified by having g$^{(2)}(0)<0.5$) out of over 100 tested across several experimental attempts.  These devices have saturation intensities from 337 to 704 kcps, which are consistently higher those from Ag-coated nanoposts.  To achieve true scalability of the system, certain material issues need to be resolved.  These include the reliable deposition of high-quality Ag films and better understanding of the chemical and electronic interactions between diamond and Ag at the interface.

The collection efficiency in our system could be further improved by the addition of an anti-reflection (AR) coating, the simplest implementation of which is a 130-nm thick layer of SiO$_2$ that can enhance transmissivity through the diamond surface by about 20\% around the peak emission of the NV center~\cite{Yeung2012, supplementary}.  Finally, optimized simulations indicate that a collection efficiency of up to 64\% as well as other beam designs are attainable~\cite{supplementary}.

\section*{Acknowledgments}
The research described in this paper was supported by the DARPA QuASAR program, and the AFOSR MURI (grant FA9550-12-1-0025).  Work by J.T.C. was supported by the Center for Excitonics, an Energy Frontier Research Center funded by the U.S. Department of Energy, Office of Science, Office of Basic Energy Services under Award Number DE-SC0001088.  B.J.M.H. gratefully acknowledges support from HQOC.  The devices were fabricated at the Center for Nanoscale Systems at Harvard University.  We thank D. Twitchen and M. Markham from Element 6 for helpful discussions and for providing samples.  We thank E. Macomber, Y. Lu, and J. Deng for their assistance with the tools.  We also thank M. Kats, D. Floyd, T. Liu, S. Cui, E. Hu, and M. Grinolds for experimental support and helpful discussions.


\newpage

\section*{Supplementary Information}
\subsection{Anti-reflection coating}
We designed an anti-reflection (AR) coating to reduce the reflection at the diamond-air interface over the critical cone (25$^{\circ}$).  For diamond, SiO$_2$ provides an ideal coating material.  We modeled different coating thicknesses using FDTD, using a broadband point source below the diamond-air interface and averaging the transmission over the acceptance cone, and found an optimal thickness of around 130 nm.  As shown in Fig.~\ref{suppfigure4}, $\sim$ 96\% of light can be transmitted at the peak wavelength of the NV center emission, which is about 20\% higher than the uncoated case ($\sim$73\%).

We deposited 130 nm of SiO$_2$ on a diamond sample with silver-coated nanoposts by sputtering and measured the saturation intensities of the same NV center before and after SiO$_2$ deposition (Fig.~\ref{suppfigure5}).  The saturation intensity and power are 138 kcps and 1.1 mW for the uncoated sample, and 151 kcps and 0.93 mW after applying the AR coating, representing about 10\% and 18\% improvements in collection and pump efficiencies, respectively.  The discrepancy with analytical prediction might be attributed to the fact that our calculation did not take into account the spectrum of the NV center, while the measured values in experiment have been spectrally integrated.  Moreover, there might be a difference between the deposited thicknesses on diamond and silicon wafer that we used to calibrate the deposition rate.

\makeatletter 
\renewcommand{\thefigure}{S\@arabic\c@figure} 
\begin{figure}[H]
  \centering
  \includegraphics[width=\textwidth]{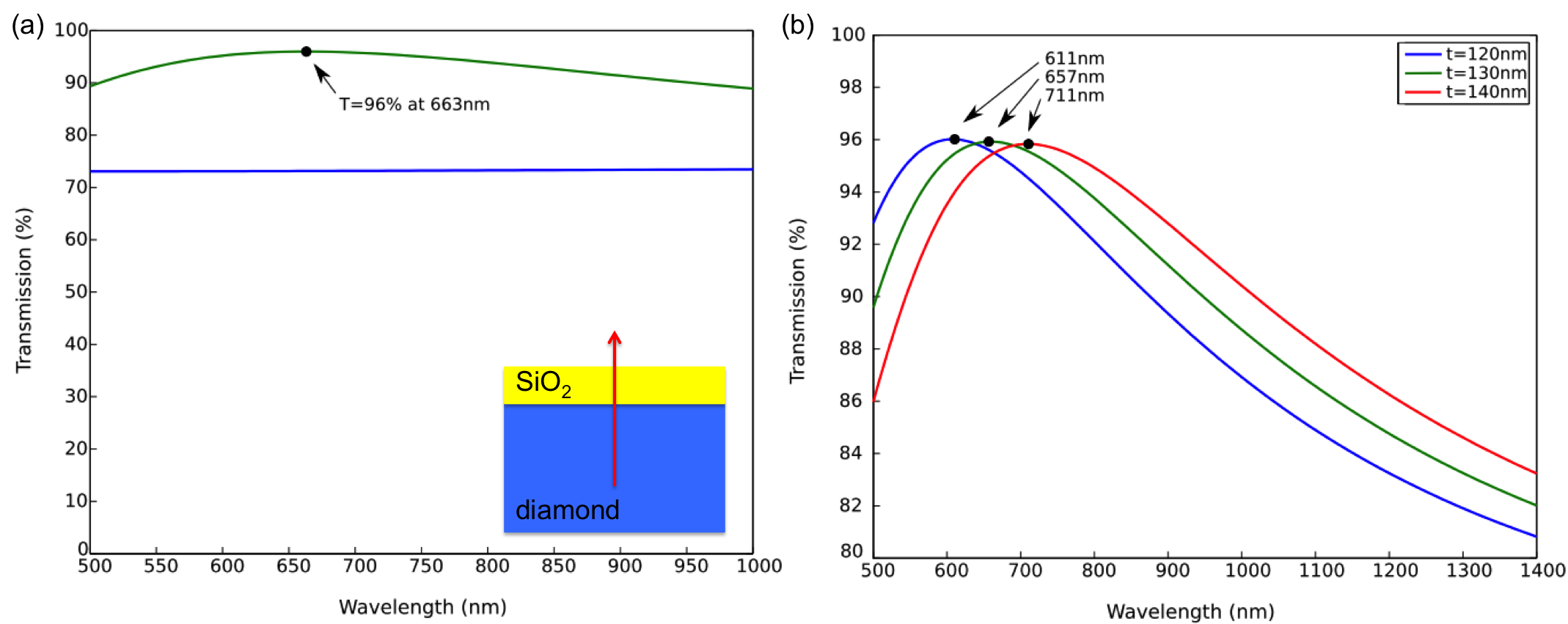}
\caption{(a) Calculated transmission spectrum of a diamond-air interface with a 130-nm thick SiO$_2$ AR coating on diamond (green) and without any coating (blue). (b) Transmission spectra for different coating thicknesses.}
\label{suppfigure4}
\end{figure}

\begin{figure}[H]
  \centering
  \includegraphics[width=5.5in]{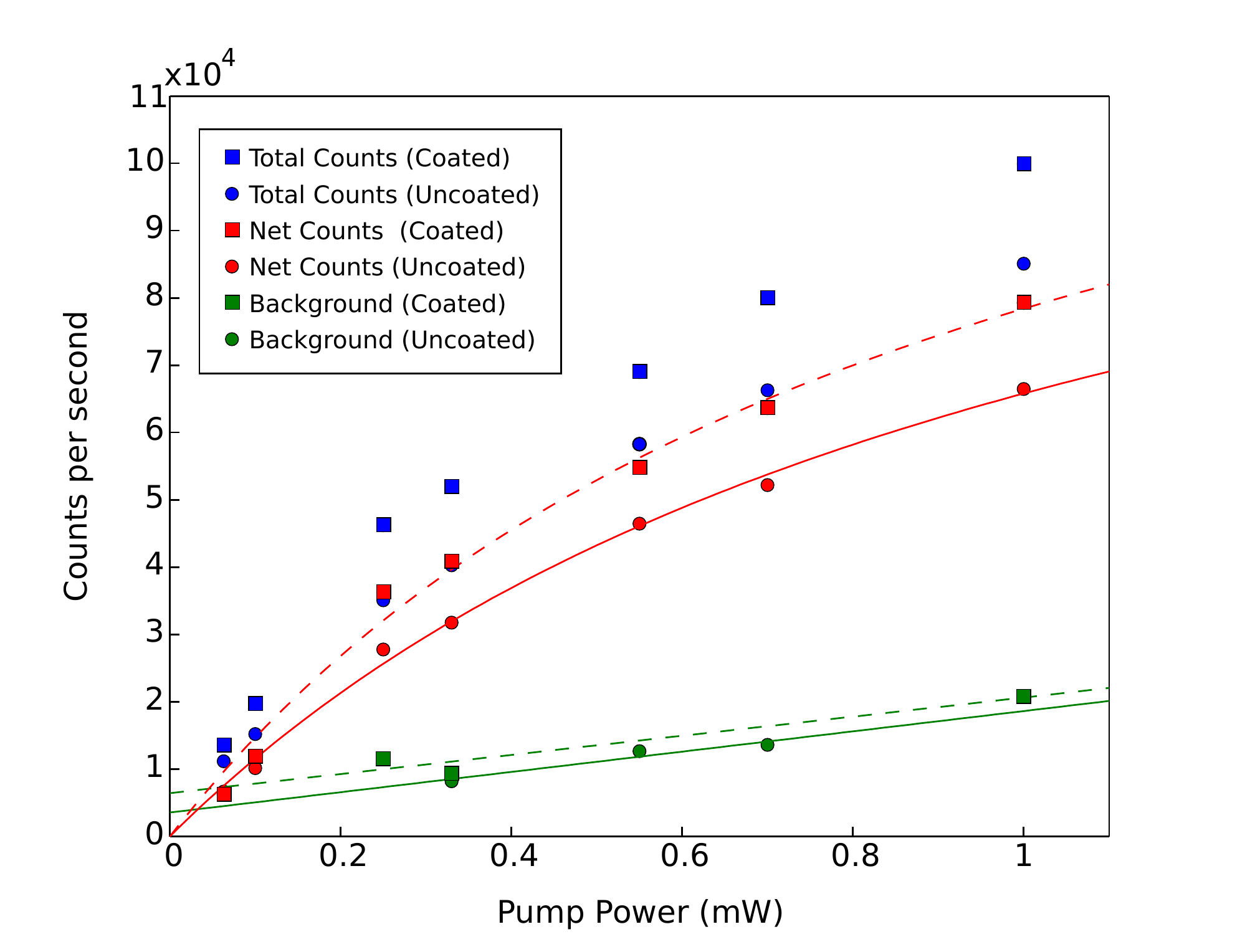}
\caption{Saturation curves of the same (silver-coated nanopost) device with (dotted red line) and without (solid red line) AR coating.}
\label{suppfigure5}
\end{figure}

\subsection{Optimized designs for better collection}
\begin{figure}[H]
  \centering
  \includegraphics[width=4.5in]{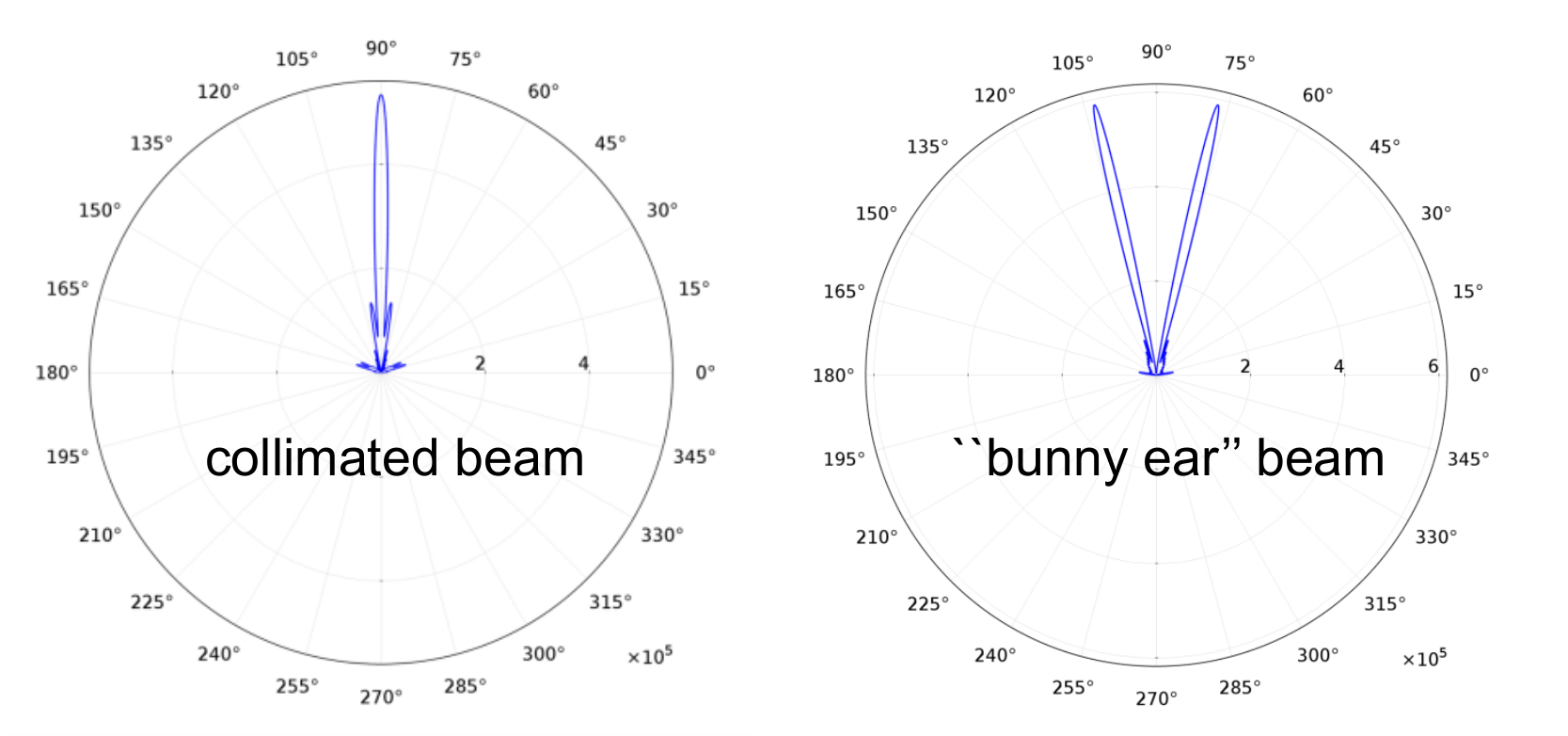}
\caption{Simulated angular emission of optimized designs.}
\label{suppfigure7}
\end{figure}

We optimized our device design and obtained a collection efficiency of up to 64\% for a strongly collimated beam (Fig.~\ref{suppfigure7}).  This revised design requires 15 layers of grooves with varying periodicities, and was generated using the optimization module on COMSOL to maximize the power emitted into a narrow angular cone ($\sim10^\circ$) normal to the sample plane in the far-field.  By maximizing the power emitted into a narrow range of angles around a specific polar angle (at 15$^\circ$ away from the axis, for example), we obtained a ``bunny ear'' beam shape with a collection efficiency of 46\% (Fig.~\ref{suppfigure7}).  Such a distinctive angular profile might facilitate the demonstration of beam shaping in fourier plane imaging on our moderate N.A. setup.  For practical applications, the general geometry and design approach can be used to generate on-chip optical components such as collimators and beamsplitters that are integrated with single emitters.

\subsection{Effects of tapering}
\begin{figure}[H]
  \centering
  \includegraphics[width=4.5in]{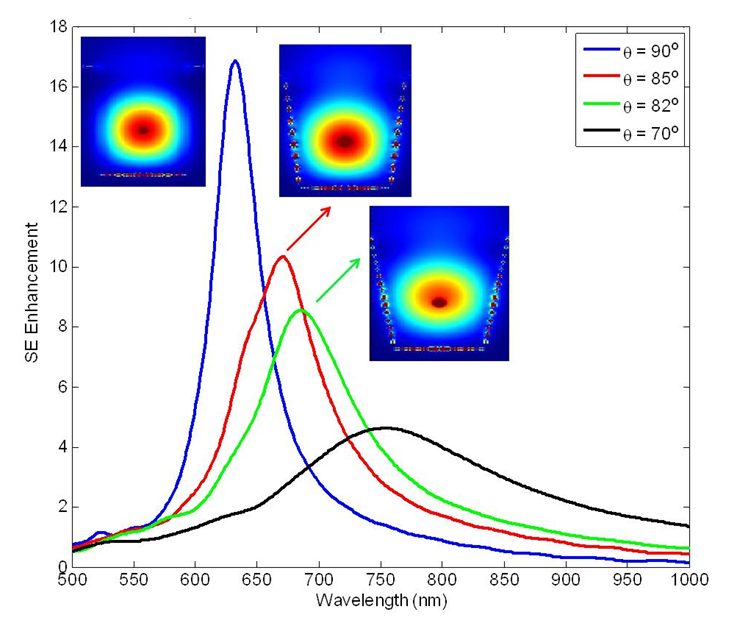}
\caption{Simulated plot of spontaneous emission (SE) as a function of wavelength for different sidewall angles.  In all the cases, the radius at the top (smaller) facet of the nanopost is the same.  Insets: the electric field intensity profiles.  These were taken before the dipole source has decayed completely, so there's still an intense spot in the location of the dipole.}
\label{suppfigure6}
\end{figure}

We investigated the effect of having slanted sidewalls in our nanoposts that resulted from our reactive ion etching process, by varying the sidewall angle (from 90$^\circ$ to 70$^\circ$) and calculating the spontaneous emission enhancement spectra (Fig.~\ref{suppfigure6}).  The resonant wavelengths red-shift with increasing taper, and the enhancement factor decreases due to reduction in the confinement of the electric field as the nanopost becomes wider.  

\subsection{Modeling}
\begin{figure}[H]
  \centering
  \includegraphics[width=4in]{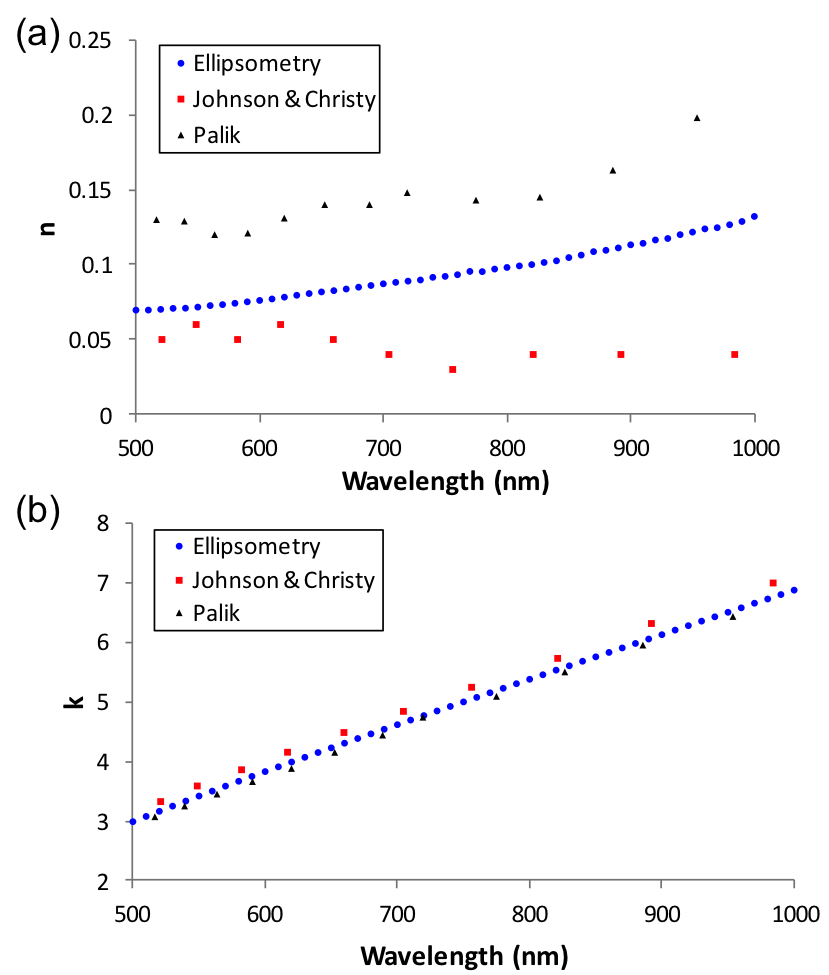}
\caption{(a) $n$ and (b) $k$ obtained from spectroscopic elllipsometry on our sputtered silver film, along with reported values from literature.}
\label{suppfigure1}
\end{figure}

The material parameters used for our simulations are as follows: the refractive index ($n$) of diamond and SiO$_2$ are 2.4 and 1.46, respectively.  Both are assumed to be lossless over the wavelength range of interest.  We performed spectroscopic ellipsometry on sputtered silver films and determined their $n$ and extinction coefficients ($k$) (Fig.~\ref{suppfigure1}).  Our data are mostly consistent with values reported in literature (Palik~\cite{Palik} and Johnson \& Christy~\cite{JohnsonChristy}) and vary more smoothly with frequency.  This facilitates the proper construction of an analytic material model that is required for our FDTD simulation.
 
As seen in Fig.~\ref{suppfigure1}, silver has very small values of $n$, which when placed next to materials with high dielectric constants (such as diamond), can lead to spurious solutions due to the large mismatch in $n$.  Consequently, the simulations require very fine (sub-nm) meshing, which greatly increase the computation time.  We found that by adding a very thin ($<10$ nm) spacer layer of a lower dielectric constant (such as Al$_2$O$_3$ ($n = 1.67$)) surrounding the diamond nanopost, the solution is able to converge at larger meshing (2 nm) and only slightly shifts the resonant response.  

\subsection{Background extrapolation from g$^{(2)}$ measurements}
\begin{figure}[H]
  \centering
  \includegraphics[width=\textwidth]{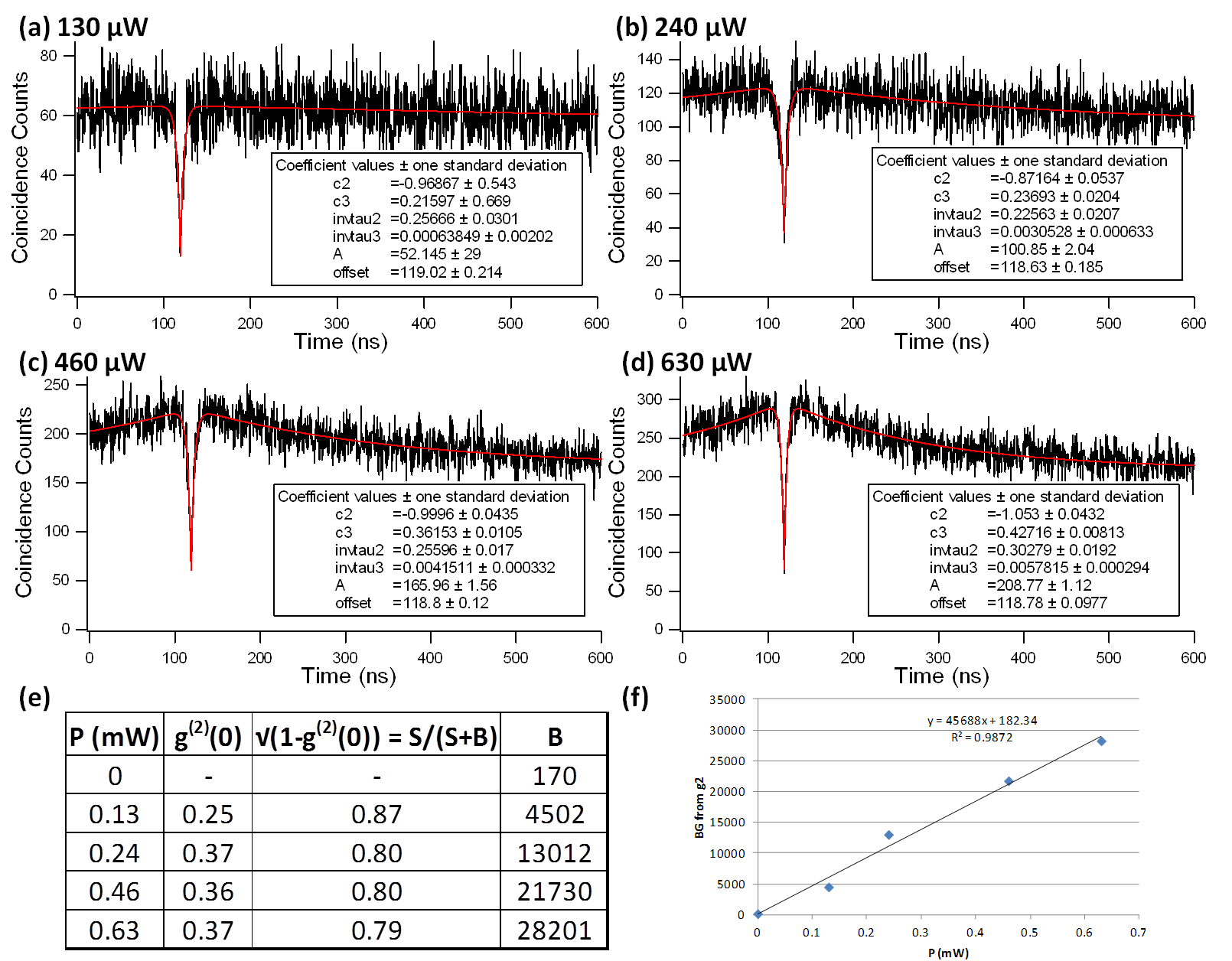}
\caption{(a)-(d) Raw g$^{(2)}$ data on a grating device (shown in Fig. 4 in the main text) taken at different pump powers with fits (red).  (e) Tabulated values for g$^{(2)}$ and background at different powers.  $B$ at zero power refers to the total dark counts from the APD.  (d) Plot of background as a function of pump, with a linear fit.}
\label{suppfigure2}
\end{figure}

\begin{figure}[H]
  \centering
  \includegraphics[width=\textwidth]{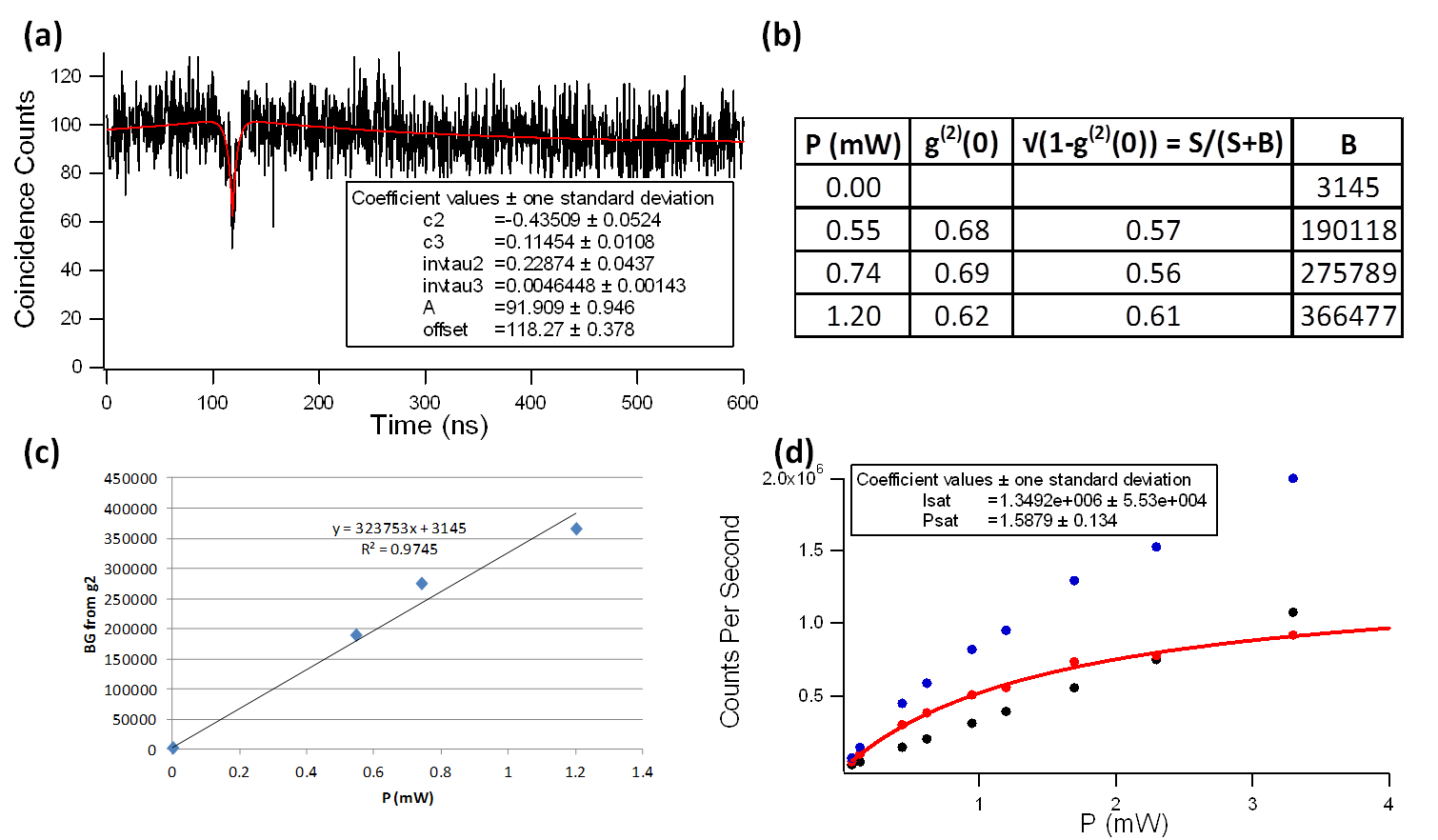}
\caption{(a) Raw g$^{(2)}$ data for the same grating device taken with a multi-mode fiber.  (b) Table of g$^{(2)}$(0) and background values.  The difference in dark counts from Fig.~\ref{suppfigure2} is due to the increased scattering of stray light into the fiber.  (c) Plot of background vs. pump. (d) Saturation curve in which blue represents the total count rate from the device; black is the interpolated background; red is the background-subtracted signal (NV fluorescence).}
\label{suppfigure3}
\end{figure}

Here we provide details of the background extrapolation method~\cite{Beveratos2001, Hausmann2011} we used for all of our saturation measurements.  We acquired g$^{(2)}$ at different pump powers and fitted the raw data to the model g$^{(2)}(\tau) = A (1+c_2e^{-|\tau-offset|\Gamma_2}+c_3e^{-|\tau-offset|\Gamma_3})$.  We then obtained values at zero delay, which along with the total count rates during the g$^{(2)}$ acquisition, give us the background values.  More details are provided in Figs.~\ref{suppfigure2} and \ref{suppfigure3}.

\end{document}